\documentclass[12pt]{article}
\usepackage{kantlipsum} 
\usepackage[utf8]{inputenc}
\pdfoutput=1
\title{  \large{Education Policy and Intergenerational Educational Persistence: Evidence from rural Benin\footnote{ I am grateful to Marc Henry, Hirano Keisuke and Gechter Michael David for the helpful comments and suggestions. I also thank Lewis McLean and Manjunath Amrutha for their useful suggestions.}} }
\author{Christelle Zozoungbo\footnote{Pennsylvania State University: Department of Economics (cfz5115@psu.edu)}}
\date{July, 2023}
\usepackage{subfigure}
\usepackage[round]{natbib}
\usepackage{graphicx}
\usepackage{amsmath}
\usepackage[a4paper, total={7in, 9.5in}]{geometry}
\usepackage{hyperref}

\hypersetup{
  colorlinks,
  citecolor=blue,
  linkcolor=blue,
  urlcolor=blue}
  
\usepackage[dvipsnames]{xcolor}
\usepackage{setspace}
\usepackage{comment}
\usepackage{bbm}
\usepackage{float}
\usepackage{tablefootnote}
 \usepackage{setspace}

\usepackage{tikz}
\begin{document}

    \maketitle
    \begin{abstract}
    \small{
 \setstretch{1}
This paper employs a nonlinear difference-in-differences approach to empirically examine the Maximally Maintained Inequality (MMI) hypothesis in rural Benin. The findings of this study confirm the MMI hypothesis. In particular, it is observed that when 76\% of educated parents choose to educate their daughters in the absence of educational programs, in contrast to only 37\% among non-educated parents, the average impact of tuition fee subsidy on enrollment probability in primary schools stands at 3.8\% for non-educated households and 0.27\% for educated households. Conversely, in cases where only 27\% of educated parents decide to educate their daughters without education programs, the average effect of tuition fee waivers on enrollment probability in primary schools increases to 19.64\% for non-educated households and 24\% for educated households.
From the analysis of household education decisions which is influenced by preference for education and budget constraint, three key conclusions emerge to explain mechanism behind the MMI. Firstly, when the income advantage of educated households compared to  non-educated households is significantly high, irrespective of the level of their preference advantage, reducing the financial cost of education induces a greater shift in education decisions among non-educated households. Secondly, in situations where educated households do not possess an income advantage relative to non-educated households, the reduction in education-related financial costs leads to a more pronounced change in education decisions among educated households. Lastly, for the low income advantage of educated households, as the income advantage of educated households increases, non-educated households respond more to education policy compared to educated parents, if the preference advantage of educated households is relatively smaller.

}
    \end{abstract}



%\newpage
%\tableofcontents
\newpage

\section{Introduction}
\paragraph{}
Education embodies not only a facet of development but also a conduit for its realization. Within the sphere of human capital formation, education stands as a pivotal element, alongside health, as highlighted by \cite{de2015development}. In the pursuit of development, numerous developing nations have undergone substantial expansions in education, encompassing initiatives like erecting schools and reducing school fees. These programs are primarily aimed at curbing the financial burden of education, thereby fostering heightened demand for schooling within lower socioeconomic strata.
Despite generally yielding positive outcomes—both in terms of enrollment rates and subsequent adult earnings, as documented by \cite{duflo2001schooling}, \cite{chen2013impact}, and \cite{ashraf2020bride}—these programs have engendered a contentious debate regarding their impact on intergenerational educational persistence. The determination of households to invest in education hinges on an array of factors, including anticipated returns, opportunity costs, and the inherent value attributed to education. As argued by \cite{de2015development}, parents with limited incomes exhibit diminished investments in their children's education due to lower expected returns and intrinsic valuation of education, coupled with heightened opportunity costs. Consequently, this phenomenon becomes a conduit for the perpetuation of poverty from one generation to the next.
Furthermore, the threshold for downward mobility—defined as a lower educational attainment relative to parental levels, as expounded by \cite{boudon1974education}—is notably lower for students whose parents possess limited education or income. In contrast, students hailing from more educated and financially prosperous families are inherently better equipped, both cognitively and non-cognitively, to capitalize on the emerging educational opportunities presented by these programs.

Consequently, education programs might increase inequality of educational opportunities (IEO) rather than decrease it. In essence, these education programs exhibit an inherently inequitable impact. The foundation of this conclusion is built upon the premise of the Maximally Maintained Inequality (MMI) hypothesis, positing the coexistence of both educated and non-educated parents who opt not to enroll their children in school. Specifically, the benefits of education programs tends to favor more educated parents as long as a substantial proportion of these parents refrains from enrolling their children due to financial constraints or the perceived benefits of education failing to outweigh its opportunity costs. In such a scenario, educational initiatives targeting primary education, for instance, could potentially yield either an increase or no discernible alteration in IEO. Conversely, within a population where children from educated or high-income families have already achieved a saturation point at a particular educational level, the introduction of education programs is more likely to result in a reduction of IEO, as highlighted by the findings of Raftery and Hout (1993).
In essence, the impact of education programs is multifaceted and varies significantly across different family groups, underscoring the heterogeneous nature of their effects.

This paper delves into an examination of the heterogeneous impact arising from an elementary education subsidy, situated within a context characterized by two distinctive attributes. Firstly, intergenerational educational persistence is evident, as shown by the higher proportion of children who have received at least one year of education within households where at least one parent possesses a minimum of one year of education. Secondly, a notable dissimilarity exists between educated parents and their non-educated counterparts regarding the presence of non-educated children; the absence of education programs sees a relatively small percentage (24\%) of educated parents with non-educated children, compared to their non-educated counterparts (62\%). The specific focus of this analysis is to evaluate the efficacy of an education subsidy program within rural areas of Benin, particularly its impact on girls' education. In 2003, the Beninese government took the significant step of waiving tuition fees and any associated parental contributions for girls enrolled in public primary schools across rural regions. This policy was subsequently extended to encompass the entire country in 2006. Drawing upon data sourced from the 2013 Population and Habitation Census in Benin, this study aims to assess the effectiveness of the 2003 decision to introduce free elementary education. To achieve this, a meticulous examination is conducted to understand how the policy influenced the probability of enrollment in and completion of primary school among girls in rural settings. This evaluation is undertaken within the framework of the education level of the household's head and the household's income level, delineating the nuanced variations in the impact of waived tuition fees across distinct family groups.

I use as identification strategy the nonlinear difference in differences \citep{ai2003interaction} estimation method. I exploit the fact that my data contains children from every cohort from 1953 to 2000, to separate households with children exposed to the reform from those who were not, based on the age in 2003 of children in the data. The average age of children in elementary school is between 6 and 11 years old in Benin. And the official age to enroll in primary 1 is 6 years old. However, it is common to enroll 7 or 8 years old children in primary 1, especially in rural areas, which explains why primary school gross enrollment rate\footnote{The elementary school gross enrollment rate is the ratio between the number children enrolled in elementary school and the number of children at official age to enroll in elementary school (which is between 6 and 11 years old).} happens to be larger than 100\% sometimes (Report of UNICEF Benin 2017, Chapter 4). In addition, our data shows that there is an insignificant drop out rate in primary school in Benin (Figure \ref{hist_educ}). This implies that enrollment and completion of primary school are substitutes. I, therefore,  focus on enrollment in primary 1 as outcome variable in my analysis.
It follows that, children older than 8 years old in 2003 were not exposed to the reform. In other words cohorts born in 1985 to 1994 are cohorts for which education decision was made before the policy. While cohorts born in 1995 to 2000 are cohorts whose education decision happens after the policy. 

I consider the household as unit of observation, and get a pseudo panel data where households education decision is observed multiple times through their children. My treatment group constitutes of households with daughters living in rural areas. 
Using households with sons living in rural areas as control, the nonlinear difference in differences estimate shows that households with non-educated head of household uniformly benefit more from the reform. In other words the empirical distribution of the treatment effect on the treated for households with non-educated head of household first order stochastically dominates the one for households with educated head of household. This conclusion, which seems to be in conflict with the theory on education expansion in the context of intergenerational educational persistence, is explained by the structure of the targeted population in this particular context. Indeed, an estimate of the conditional probability of a daughter to be educated given she is from a household with educated head of household ($\approx 76\%$) is more than twice the estimated conditional probability of a daughter to be educated given she is from a household with non-educated head of household ($\approx 38\%$) before the policy in rural areas (Figure \ref{fig3}). In other words daughters from educated household are more likely to be educated even in the absence of a decrease in cost of schooling by an education policy. That is an indicator of intergenerational educational persistence. However, before the policy, among educated parents only $24\%$ have non-educated daughters. This signals that educated parents need less financial incentive to enroll their children in school than non-educated parents do.

In order to analyze the counterfactual result where there is a relatively higher proportion of educated parents with non-educated children, I include in the regression an interaction between the effect of the education policy and a variable indicating if the household has at least one non-educated child before the policy. This gives a completely different structure of the population before the policy, where we still have intergenerational educational persistence but less strong in the sense that among educated parents $73\%$ have non-educated daughters compared to $86\%$ for non-educated parents. The same analysis as before now results in educated parents having uniformly higher effect from the policy.

I finally consider a household education decision problem in order to investigate how preference for education\footnote{Preference for education is reflected in the differences in education decisions and determined by the difference between expected return/intrinsic value for education and opportunity cost of education.} interact with budget constraint to result in a given education decision by households. How preferences and budget constraint are affected by the elimination of financial cost of education and how this translate into change in education decision. I derive conditions on distributions of income and value for education under which non-educated households benefit more from education policy.

The subsequent sections of this paper unfold as follows: Section 2 offers an extensive review of pertinent literature, juxtaposed with the distinctive contribution of this study. Presenting the contextual background of the reform, Section 3 elucidates the environment in which the policy unfolds. Proceeding onward, Section 4 provides an in-depth exposition of the data and variables employed in the analysis. Section 5 is dedicated to outlining the strategic identification strategies implemented for estimating the policy's effect. Transitioning into Section 6, the estimation results are comprehensively presented and discussed. In section 7 I present the household's education decision problem and section 8 concludes the paper.

\section{Literature Review}
This paper contributes to the literature on impact evaluation of education programs in developing countries (Duflo (2001), Chen et al (2013), Ashraf et al. (2020)). \cite{duflo2001schooling}, using evidence from school construction program in Indonesia, has shown that the construction of primary schools led to an increase in schooling attainment and earnings for a sample of men. Along a similar line  \cite{ashraf2020bride} have shown that the same program has also a positive effect for a sample of women when they account for a particular cultural practice (The bride price). \cite{chen2013impact} found that a tuition fees subsidy program has a positive effect on the math achievement of poor junior high school students in rural China. \cite{duflo2021impact} show that free secondary education in Ghana through scholarship grants has a positive effect on education attainment, knowledge and skills acquisition and other social factors such as health behaviors and female fertility. This paper contribute to that literature by showing that elimination of financial cost of primary school for girls in rural Benin increases their enrollment probability.

This paper also contributes to the literature on  intergenerational mobility in terms of human capital formation and educational attainment (\cite{boudon1974education}, \cite{breen1997explaining}, \cite{checchi2008intergenerational},  \cite{torche2019educational}, \cite{chusseau2013education}). \cite{checchi2008intergenerational} have shown that the persistence inequality in college education enrollment in Italy is mainly due to differential liquidity constraint and risk aversion between parents with and without college education. \cite{chusseau2013education}, have demonstrated that one potential explanation for the rise of under-education trap\footnote{Under education is the situation in which non educated families remain non educated from one generation to the next} is the fixed cost of education. They show that fixed cost of education is a sufficient condition for under education traps to emerge. More specifically, they have shown that under fixed education cost, individuals whose parents' human capital is below a certain positive threshold choose not to educate themselves. This yields a human capital of zero for those individuals, implying that all following descendants will not be educated. Therefore a decrease in the cost of education will lead to a decline of the threshold of human capital below which parents do not educate their children. This is because decrease in cost of schooling makes education affordable to parents with low human capital. But more generally the cost is lower for everyone, both low and high human capital parents. Therefore both low and high human capital parents are inclined to benefit from it.  In the present paper we investigate how a policy that decrease the cost of education increase the number of educated individuals with low human capital parents compared to those with high human capital parents.

\section{Context of the reform and its implementation}

\paragraph{}
A number of developing countries have invested substantial efforts in policies and programs to make education, at least elementary education, free to everyone. This in order to stay in line with the article 26 of the Universal Declaration of Human Rights, according to which \textit{``everyone has the right to education and that elementary education shall be compulsory"}. 
The republic of Benin is no exception, as clearly stated in the article 13 of its constitution (December 1990): \textit{the government shall provide education to everyone through public schools, and elementary education is compulsory}. Therefore the government's objective is to assure progressively that public education is free for everyone with priority given to disadvantaged and unprivileged population. 

In order to achieve that objective in addition to the second goal of the Millennium Development Goals (MDGs), the government of Benin took some concrete actions regarding elementary education in 2003 and 2006. Specifically, in addition to the decision of free elementary education for girls in rural areas in 2003, the government decided in 2006 that elementary education is free for everyone in Benin.  These decisions are made by the law  n°2003-17 of November 2003 concerning national education, modified by the law  n°2005-33 of October 2005 which defined a new legal framework for public education. The law  n°2003-17 highlights in its article 03 that priority should be given to girls, people in difficult situation and vulnerable groups.

Even though subsidies and additional classrooms construction programs have been implemented in order to sustain the decision (European Union Report, 2019), the implementation of the free elementary education in Benin does not rely on any administrative or legislative act. There is no decree issued by the government, much less a law that gives legality to this decision. Although this decision has solid legal foundations, reference should be made to October 14, 2006 Ministers Council's statements, which decreed it to give it a legal character (Report of Benin Ministry of Finance, 2012). 
In this paper we will analyze the decision of free elementary education for girls in rural areas in 2003.

\section{Data and Variables}
The main data used in this paper is from the 2013 Population and Habitation Census in Benin (a sample of 10\%). The data is available on IPUMS website. It provides information at the household level such as households' characteristics, assets and utilities owned. It also provides information at the individual level. The individual level variables include education attainment, gender, age, marital status, religion and ethnicity of each member of the household for around 180,000 households from rural and urban areas. A variable in the data giving the relationship of each member with the head of household allows me to match children to one of their parents in the sample. I am able to construct a sample of children and one of their parents -- father if the head of household is a man and mother if the head of household is a woman. In order to guarantee that individuals in the sample have not been exposed to the 2006 policy, I reduced the sample to individuals of at least 13 years old in 2013 (the time of the census). Using indicator variables for household assets and utility services (type of toilet, type of flooring, television and refrigerator ownership ...) I constructed a wealth index for each household using principal components analysis procedure. 
I will focus on elementary education attainment for the analysis, giving that the policy I am analyzing is at elementary education level. Another reason why I decided to focus on elementary education is that my data is only 10 years after the implementation of the policy, meaning that children that were exposed would have not even finished junior/senior high school.

Define $Y_{it}$ as the elementary education attainment of individual of age (or cohort) $t$ from household $i$. $Y_{it} = Y_{it}^{\star} \times  \mathbbm{1}{ \{ 0 < Y_{it}^{\star} <6\} } + 6 \times \mathbbm{1}{ \{ Y_{it}^{\star} \geq 6\} }$, where $Y_{it}^{\star}$ is the number of years of schooling.

\subsection{Construction of the Variable ``Household Wealth Index (HWI)"}
I construct a household wealth index that I will use as measure of household's wealth. I use the same principle used for the construction of the DHS Wealth Index  \citep{rutstein2004dhs}. I consider assets and services presented in Table \ref{table:1.1} to construct the wealth index. They are assets and services used to construct the DHS Wealth Index.
\begin{table}[h]
\caption{Assets and Services owned by the household}
\centering
\begin{tabular}{ p{15cm}}
\\
\hline
\hline
Type of flooring (earth, cement, wood, tile)\\

Water supply (piped inside/outside dwelling, public piped water, no piped water)\\

Sanitation facilities (no toilet, flush toilet, latrine)\\

Electricity (yes/no), Television (yes/no), refrigerator (yes/no), Internet (yes/no) \\

Telephone (yes/no), Computer (yes/no), Automobile (yes/no)\\
\hline
\hline
\label{table:1.1}
\end{tabular}
\end{table}

Next, I break the variables in Table \ref{table:1.1} into indicator variables. The idea is to have a sense of ordering in wealth level based on the modalities of each variable. Figure \ref{fig:1.1} below illustrate that for the variable water supply and sanitation facilities. 

\paragraph{}

\begin{figure}[H]
    \centering
\begin{tikzpicture} 

\filldraw
  (0,0) circle (2pt) node[align=left,   below] {\textcolor{red}{Poorer -----$>$}\\ no piped water} --
  (0,0) circle (2pt) node[align=left,   above] {no toilet} --
(6,0) circle (2pt) node[align=center, below] {latrine}     -- 
(6,0) circle (2pt) node[align=center, above] {\\piped water outside dwelling/public piped water\\}     -- 
(13,0) circle (2pt) node[align=right,  above] {flush toilet}
(13,0) circle (2pt) node[align=right,  below] {\textcolor{blue}{Wealthier}\\piped water inside dwelling};
\end{tikzpicture}
    \caption{Wealth level as function of water supply and sanitation facilities}
    \label{fig:1.1}
\end{figure}
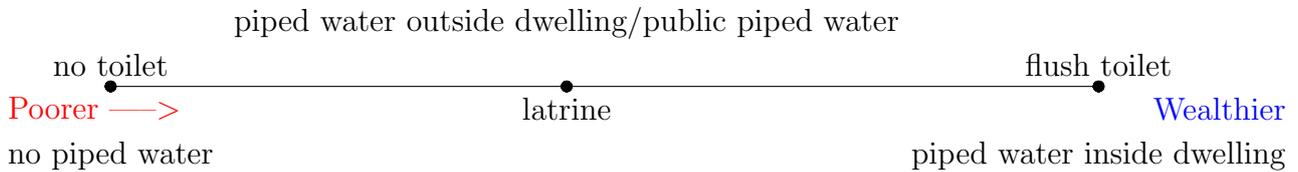

\begin{figure}[H]
 \centering
   \includegraphics [width=10cm]{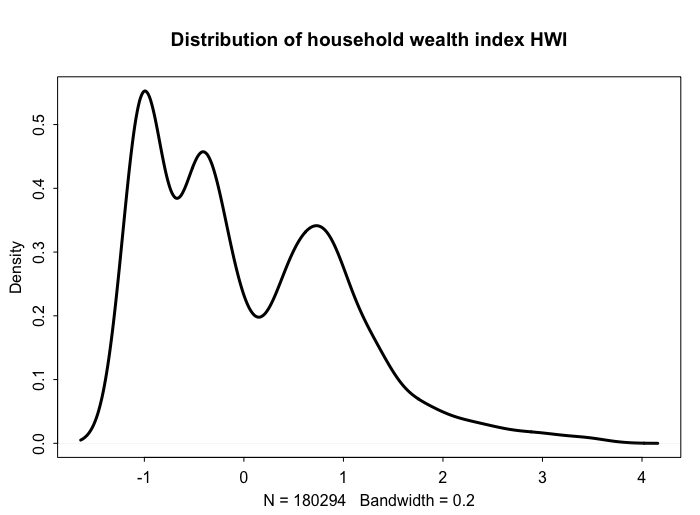}
 \caption{ Density plot of Household Wealth Index.}
   \label{fig2}
\end{figure}

\begin{figure}[H]
 \centering
   \includegraphics [width=10cm]{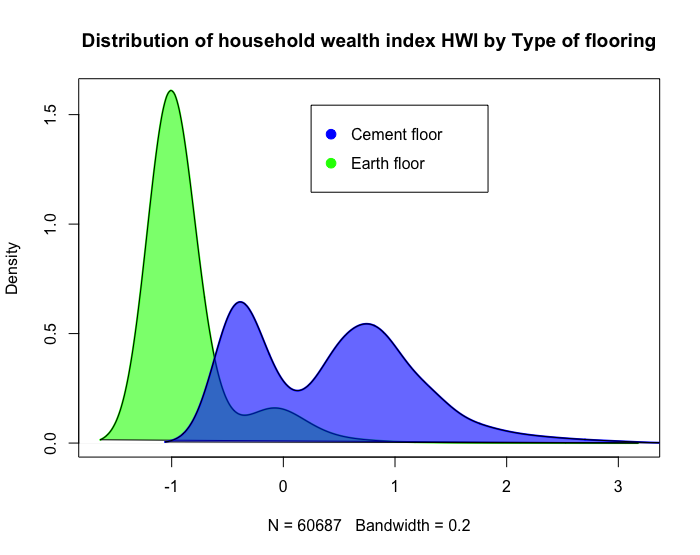}
 \caption{ Density plot of Household Wealth Index by type of flooring.}
 \label{fig3}
\end{figure}

After transforming variables into indicator variables, I use the factor analysis procedure to determine the indicator weights. The first factor produced is used as household wealth index following \cite{rutstein2004dhs}. The resulting HWI is standardized with mean 0 and variance 1. The density of HWI is presented in Figure \ref{fig2}, where we can see 2 modes, one below 0 and one above 0. This suggests that we have a classification where poor people have negative HWI, while wealthy people have positive HWI. In addition, all households with earth floor have negative HWI (Figure \ref{fig3} ). This motivates the use of binary variable for wealth index later in the analysis.

\subsection{Descriptive Statistics}

\begin{table}[h]
\caption{Descriptive Statistics \tablefootnote {The data comprises 180,294 households. Therefore we have information on 180,294 head of household (mothers or fathers). 160,280 of the members of households are biological adult and teenager children of the head of household. So, we are able to link 160,280 children to one of their parents in the data set. We refer to children here as teenager and adult children (aged between 13 and 60 years old) living in the same household as at least one of their parent. Note that $52\%$ of households in the data are located in rural areas. $24\%$ of head of households are women. The statistics provided in this Table are for sub sample of households in rural areas. The number of such households is 91317.}}

\centering
\begin{tabular}{ p{6cm} p{1.5cm} p{2cm}p{2.5cm}p{1.5cm} p{1.5cm}}
&&&&\\
\hline
\hline
Variables &Mean &Median & Standard Deviation & Min & Max\\
\hline

Head of household elementary education attainment &1.696& 0 &2.54 & 0 &6 \\

Daughters elementary education attainment  &3.15&4& 2.8 &0 & 6\\

Sons elementary education attainment &3.14&5& 2.8 &0 & 6\\

Household Wealth Index (HWI) &-0.42& -0.46&0.96& - 1.04&3.56\\

Age of the head of household &42.68 & 40 &15.61& 15 & 98\\

Number of biological children in a household&3.16 & 3 & 2.95& 0& 79\\
\hline
\hline
\label{table:1}
\end{tabular}
\end{table}
 
We can learn two things from Table \ref{table:1} which provides summary statistics for key variables in the data. First, on average, elementary years of education is higher for both sons and daughters compared to their parents. In addition, $51\%$ of children have a higher level of schooling than their parents, while only 10\% of children have lower education attainment compare to their parents. This gives a measure, though imperfect, of an upward mobility in education from the parents' generation to the children's generation. Second, the median elementary year of schooling is 4 for daughters and 5 for sons, but the average is the same for both.

\begin{figure}[H]
 \centering
 \begin{minipage}[b]{0.45\linewidth}
   \includegraphics [width=8cm]{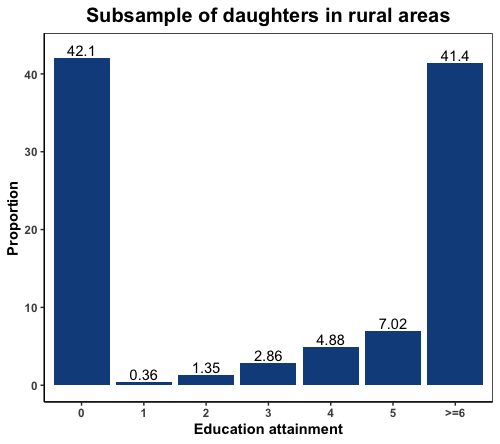}

   \label{fig:1}
 \end{minipage}
 \quad
 \begin{minipage}[b]{0.45\linewidth}
    \includegraphics [width=8cm]{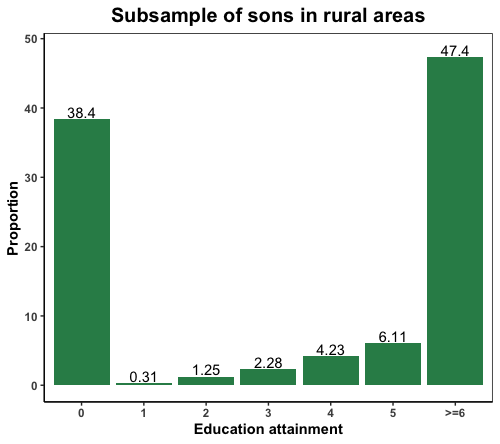}

  \label{fig:minipage2}
\end{minipage}

 \quad
 \begin{minipage}[b]{0.45\linewidth}
    \includegraphics [width=9cm]{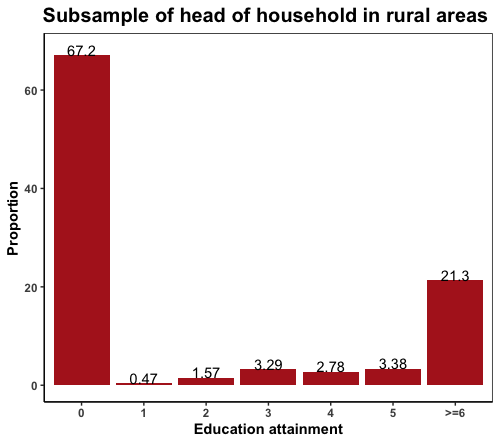}

  \label{fig:minipage3}
\end{minipage}
 \caption{ Histograms of primary school education attainment for households in rural areas.}
 \label{hist_educ}
\end{figure}

Figure \ref{hist_educ} presents the histograms of elementary education for parents and children (daughters and sons) sub-samples separately. Figure \ref{hist_educ} shows that there is an insignificant drop out rate in primary school. The highest drop out rate is 7\%, 6\% and 3\% respectively for daughters, sons and head of households, and happens in grade 5. The actual drop out rate in primary school in Benin is between 4\% and 9\% for the period 1990-2002 (World data on Education: Benin, 2006 \citep{wde}). There is almost no drop out in primary 1. That is, either people do not enroll in grade 1 at all or when they do, they complete at least grade 6. Therefore elementary school attainment is better represented as an indicator variable $Y_{it} = \mathbbm{1}{ \{ Y_{it}^{\star} > 0\} }$, which will be used in the rest of the paper.

\begin{figure}[H]
 \centering
 \begin{minipage}[b]{0.45\linewidth}
   \includegraphics [width=10cm]{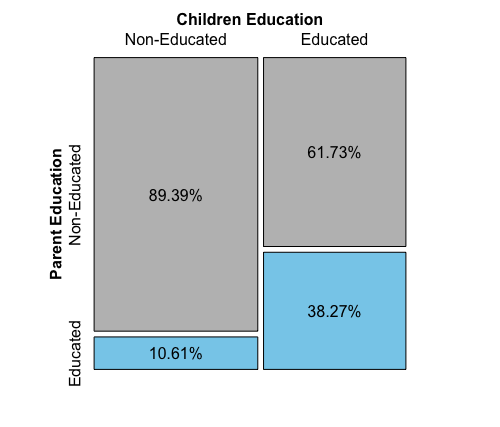}
    (a) Parents' education as function of daughters' education
   \label{f5_1}
 \end{minipage}
 \quad
 \begin{minipage}[b]{0.45\linewidth}
    \includegraphics [width=10cm]{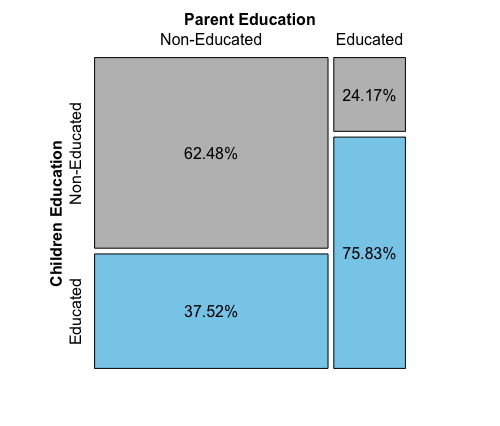}
    (b) Daughters' education as function of parents' education
  \label{f5_2}
\end{minipage}
 \caption{Education of daughters (in rural areas before the policy) and education of head of their households as function of each other.}
 \label{fig5}
\end{figure}

\begin{figure}[H]
 \centering
   \includegraphics [width=10cm]{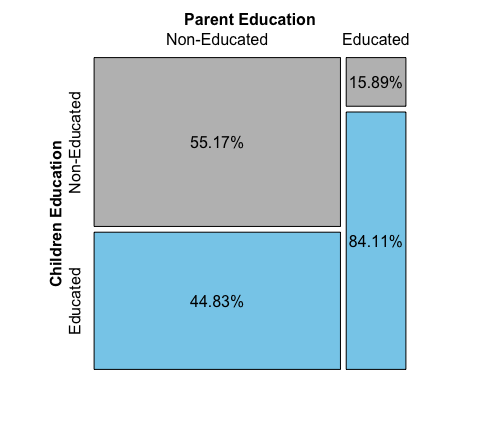}
 \caption{ Education of sons in rural as function of the education of the head of their households before the policy.}
 \label{fig4}\end{figure}

Figure \ref{fig5} shows the proportion of educated and non-educated girls in rural areas and the proportion of educated and non-educated head of their households as function of each other for the period before the policy. The first panel of Figure \ref{fig5} suggests that the target population for an education policy, which is the non-educated girls, constitutes in majority of daughters from households with non-educated head of households. Indeed, among children with no education 89\% are from households with non-educated head of households. Given that structure of the population, an education program that decreases the cost of schooling will most likely increase demand for education more from household with non-educated head of household. The first panel of Figure \ref{fig5} also shows that we have approximately the same proportion of educated and non-educated daughters in the sample. The second panel of Figure \ref{fig5} shows that we have higher proportion of households with non-educated head of household than households with educated head of household. It also shows that households with educated head of household are more likely to have daughters with at least one year of education compared to households with non-educated head of household. We observe that about two-third of children from non-educated household are non-educated, while just one-quarter of children with educated parents are non-educated. In addition as suggested by Figure \ref{fig4}, both type of household are more likely to enroll sons in school than daughters. This suggests that a household with enough resources to enroll just one child in school would prefer to enroll son if it has both son and daughter.

There are two scenarios in which the policy can have a spillover effect on sons. First, in the absence of education policy, households that can afford education for only one child choose to enroll the daughter. In this case, when the cost of education for daughters decreases, these households have additional resources to afford the education of the son. This case is ruled out by the fact that proportion of non-educated children is higher for daughters compared to sons. The second case when we can observe spillover effect from the policy is when households that can afford education for only one child choose to enroll neither son nor daughter. This case is also ruled out by the same argument. Therefore a policy that decreases the cost of schooling for daughters will not have any spillover effect on sons' education in the population.
 
\begin{figure}[H]
    \centering
   \includegraphics[scale=0.35]{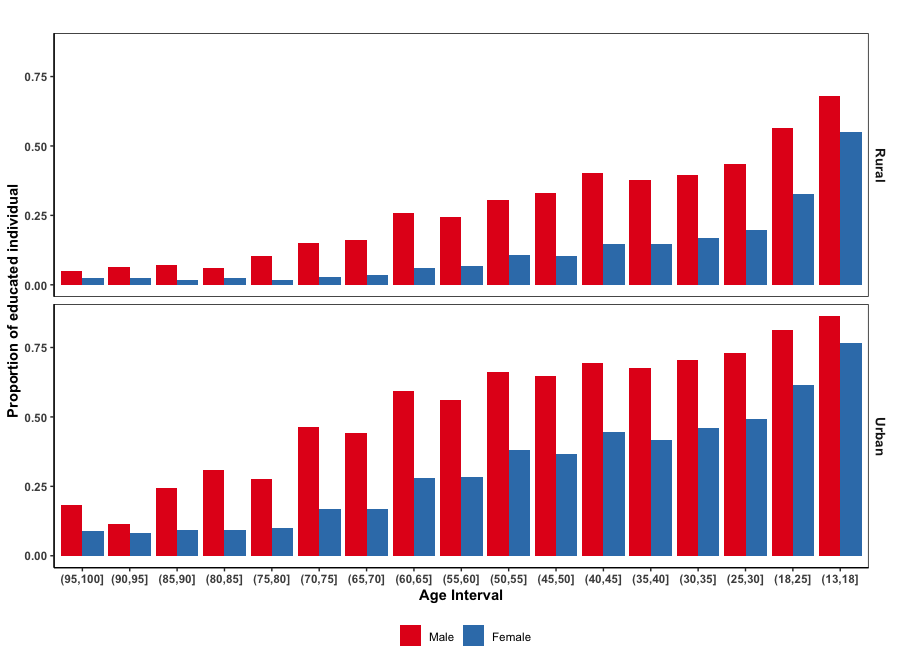}
    \caption{Proportion of educated individuals across gender and cohort.}
    \label{fig6}
\end{figure}

Figure \ref{fig6} presents the evolution of the proportion of educated individuals over time and across gender for rural and urban areas separately. The graphs rank from old generation to young. There are three major facts that can be learnt from this graph. First, the proportion of educated individuals increases over time both in rural and urban areas, and also for female and male subgroups. 
Second, the two histograms show that the fraction of educated people is higher in urban areas compared to rural areas for both gender groups and for every cohort. However, the within cohort proportion of educated males is higher compared to female both in rural and urban areas and for every cohort. This shows the inequality in education opportunities for women. Finally, for both rural and urban areas, the gap between the proportion of educated males and proportion of educated females is increasing over time, except for the youngest generation, for which the gap is smaller compare to the immediate predecessor.

\subsection{Pre- and Post- Treatment Observations}
\begin{figure}[H]
 \centering
 \begin{minipage}[b]{0.45\linewidth}
   \includegraphics [width=7.5cm]{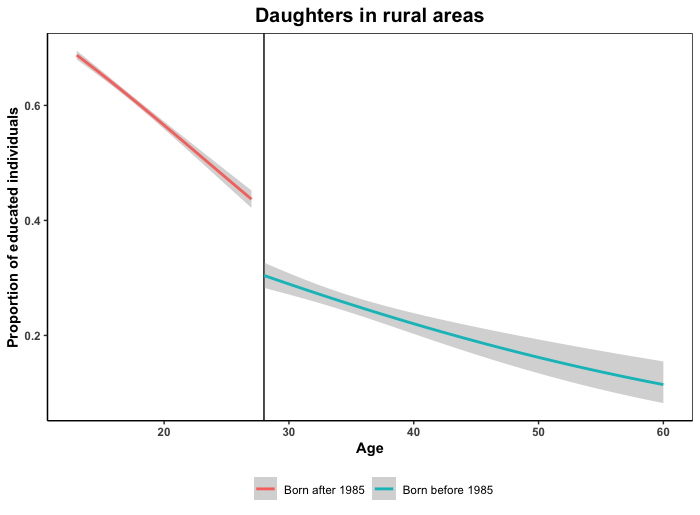}
 \end{minipage}
 \quad
 \begin{minipage}[b]{0.45\linewidth}
    \includegraphics [width=7.5cm]{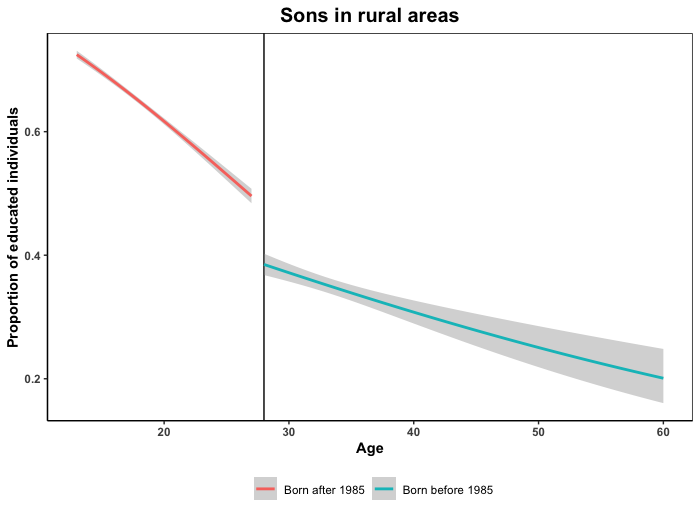}
\end{minipage}

 \quad
 \begin{minipage}[b]{0.45\linewidth}
    \includegraphics [width=7.5cm]{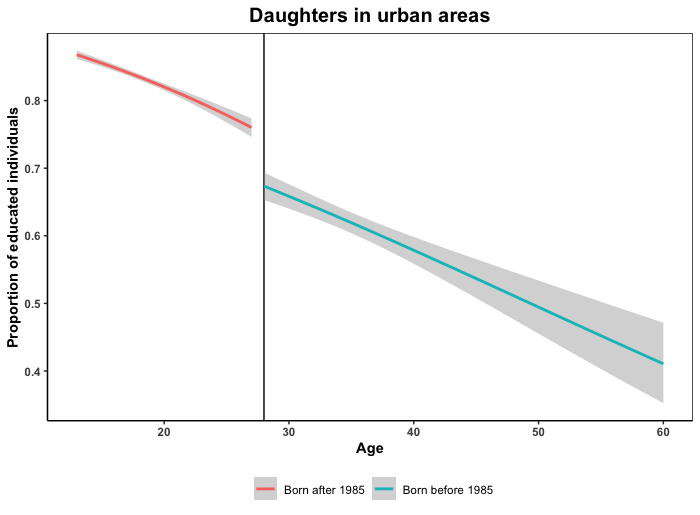}
\end{minipage}
 \quad
 \begin{minipage}[b]{0.45\linewidth}
    \includegraphics [width=7.5cm]{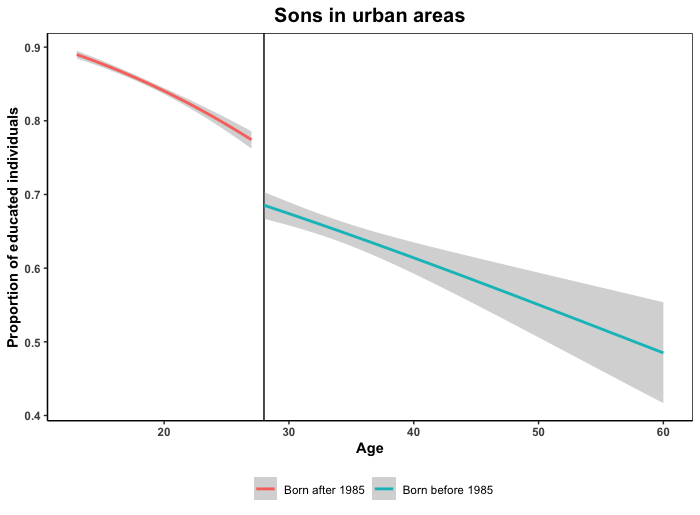}
\end{minipage}
  \caption{Proportion of educated individuals by age}
 \label{fig7}
\end{figure}

 I consider children of  18 years old or younger as post treatment observations. While children between 19 and 28 years old are considered as pre treatment observations. I have not include children  older than 28 years old in the pre treatment observations because of the structural break in the probability that an individual is educated that we observed around 1985 (Figure \ref{fig7}). A potential reason for that structural break is the December 1990 Constitution of Benin which include the first legal national measure on education. Article 13 of the constitution states that primary education shall be compulsory. Even though this measure has not been enforced, it seems to have an effect on enrollment rate.

\subsection{Sample Selection Problem}
In order to have information on the education of parents, my sample of adult children constitutes only of children living in the same household as their parents. This clearly creates a sample selection bias given that children who have succeeded both academically\footnote{Children might move to the city in order to enroll in college for example.} and financially\footnote{Children might move to the city to get a better job because they have high education.} are more likely to move out of the family house. However, this sample selection problem affects only pre-treatment observations, given that the oldest children in the post-treatment observations (18 years old) are too young to be moving out of the household even for college. In addition, the structure of households in rural Benin is more like multi-generational, where adults children have their own families in the family house with each adult child as head of its own sub household. And so children who moved out of the family house most likely migrate to the city. Analysis of the migration situation from our data suggests a relatively smaller migration to urban areas ($\approx10\%$) compared to rural areas ($\approx 19\%$)\footnote{The proportion of individuals living in the major and minor administrative units 12 months before the census is 89.87\% and rural areas and 81.76\% in urban areas.} (See Table \ref{table:6} in Appendix). Moreover, these proportions are approximately the same for men and women. Therefore bias created by educated children migrating to the city is the same both in treatment and control groups; and is therefore differentiated out in the difference in differences. Another sample selection problem is from the fact that our treatment group comprises daughters. In fact daughters who moved out of family house for example upon marriage might differ in some observable and unobservable way from daughters who still live with their parents at a seemingly old age. But this is less of a problem since my analysis is for enrollment in primary school only.

\section{Identification Strategies}
In this section, I outline the methodologies employed to assess the effectiveness of the cost reduction in girls' schooling within rural settings, while also exploring variations in its impact across different family groups. This segment will delve into the strategies and analytical tools leveraged to gauge the extent of the educational policy's effectiveness and its varying implications based on distinct household categories..
\subsection{Cross-Section Analysis with Head of Household as Unit of Observation}
First I consider a cross-section study with household as unit of observation to analyse the effect of the decrease in cost of schooling on parents decision to enroll their daughter in school. Consider as outcome variable $Y_{it} = 1\{Y^{\star}_{it} > 0\}$, the education of child of age $t$ in household $i$, where $Y^{\star}_{it}$ is the potential number of years of education. $Y_{it}$ is governed by the following equation:

\begin{equation}\label{eq:1}
    Y_{it} = \beta X_i + (\tau \Tilde{X}_i)W_{it} + \xi_i + \delta_t  + \varepsilon_{it}
\end{equation}

with:
\[\beta X_i = \beta_0 + \beta_1 \mathbbm{1} \{ \text{ Education of hh}_i > 0\}  + \beta_2 \mathbbm{1}{ \{ \text{HWI}_i \leq 0\} } + \beta_3 N_i +  \beta_4 \text{Christian}_i  +\beta_5 \text{ Muslim}_i + \]
\[  \beta_6 \text{Female}_i,\]

and
\[\tau \Tilde{X}_i = \tau_0 + \tau_1 \mathbbm{1}{\text{Education of hh}_i =0\}}+ \tau_2 \mathbbm{1}{ \{ \text{HWI}_i \leq 0\} } + \tau_3 \text{Christian}_i + \tau_4 \text{Muslim}_i + \tau_5 N_i \]
\[ +  \tau_6 \text{Female}_i\]

where, $\text{Education of hh}_i$ is the years of education of the head of the household, $HWI_i$ is the household wealth index ,  $N_i$ is the number of siblings, religion--- Christian, Muslim and others---, and $Female_i$ is the gender of the head of the household--- equal 1 if the head of household is a woman. $W_{it}$ is the treatment variable, equal one if the child is a girl in rural area and from the post-policy cohorts ($13 \leq t \leq 18$). More specifically, $W_{it} = G_{it}. T_{it}$, with $G_{it} = \mathbbm{1} \{Female_{it} = 1\}$ and $T_{it} = \mathbbm{1} \{ 13 \leq t \leq 18\}$.  $\xi_i$ is the household's unobserved characteristics and $\delta_t$ is the child age fixed effect. $\tau = (\tau_0, \tau_1, \tau_2,\tau_3,\tau_4,\tau_5)$ are our parameters of interest.

It follows from equation (1) that the proportion of educated girls in household $i$ is:
\begin{equation}\label{eq:2}
     Y_i  = \beta X_i + (\tau \Tilde{X}_i)R_i + \frac{1}{n_i^g} \sum_{t \in i} \delta_t + \mu_i
\end{equation}

where \[\mu_i = \xi_i +  \frac{1}{n_i^g} \sum_{t \in i} \varepsilon_{it}.\]
$n_i^g$ is the number of daughters in household $i$, and $R_i$ is the proportion of daughters exposed to the policy in household $i$.

Equation \ref{eq:2} identifies the treatment effect if the following assumption hold.
\begin{itemize}
    \item Assumption 5.1: $\mu_i$ and $X_i, \Tilde{X}_i$ are uncorrelated after controlling for within household average age of girls fixed effect.
\end{itemize}

By considering heads of household as units of observation we have to rely on assumption 5.1 for identification. Assumption 5.1 states that observed and unobserved household characteristics that influence education decision are uncorrelated. However, unobserved factors like preference toward sons affects the decision to enroll daughters in school, but is also correlated with parent education and wealth level.  

Therefore, we consider in the next section a nonlinear difference in differences\footnote{Given that our outcome variable is binary.} using households with sons in rural areas as control group.

\subsection{Nonlinear Difference in Differences with Sons in Rural Areas as Control group}
In this section, we consider a nonlinear difference in differences (DiD) to identify the heterogeneous treatment effect on the treated. The identification assumption in a nonlinear DiD is the parallel trend assumption on the unobserved latent linear index instead of the observed discrete outcome variable. Consider a binary choice model with latent variable
\[Y^{\star}_{it} = \alpha G_{it} + \beta T_{it} + (\gamma {X}_i)W_{it}   + \theta X_i +  \varepsilon_{it}, \]
where  $Y_{it} = 1\{Y_{it}^{\star} > 0\}$, $G_{it} = 1\{Rural_i = 1 \text{ } \& \text{ } Female_{it} = 1\}$, $T_{it} = 1\{ 13 \leq t \leq 18 \}$, $W_{it} = G_{it}T_{it}$ and ($X_i$, $\Tilde{X}_i$) is as defined in section 5.1. We assume that $\varepsilon_{it}$ has a standard normal distribution, so that we have a probit model. Therefore:
\begin{equation}\label{eq:5}
    E[Y|T,G,X] = \Phi(\alpha G + \beta T + {X} W \gamma+ X \theta),
\end{equation}

The treatment effect on the treated has the following expression \citep{puhani2012treatment}\footnote{Proof of equation 4 is appendix 10.4}: 
\begin{equation}\label{eq6}
    \tau(G=1, T=1, X) = \Phi(\alpha + \beta +  {X}W \gamma+ X \theta) - \Phi(\alpha + \beta  + X \theta)
\end{equation}
The average treatment effect on the treated is given by:
\begin{equation}
    \tau = E_X[\tau(G=1, T=1, X_i]
\end{equation}
Let $\kappa = (\alpha, \beta, \gamma, \theta)$. $\kappa$ is identified from equation \ref{eq:5} and $\tau$ is identified from equation \ref{eq6}.

\subsection{Estimation and Inference}
Consistent estimator $\Hat{\kappa}$ of $\kappa$ is obtained from a probit regression of $Y$ on $G$, $T$, $X$ and $X W$. It follows from Continuous Mapping Theorem (CMT), that $\Hat{\tau}(G=1, T=1, X)$ is a consistent estimator of $\tau(G=1, T=1, X)$. With,
\[ \hat{\tau}(G=1, T=1, X) = \Phi(\hat{\alpha} + \hat{\beta} + X W \hat{\gamma}+ X \hat{\theta}) - \Phi(\hat{\alpha} + \hat{\beta} + X \hat{\theta})\]

\begin{itemize}
    \item Assumption 5.2: We assume homoskedasticity, independence between errors and regressors, and unit error variance.
\end{itemize}
Under assumption 5.2, the asymptotic distribution of $\kappa$ is given by:

\[ \sqrt{n}(\kappa - \hat{\kappa}) \xrightarrow{d} N(0, \Sigma), \]
where,
\[ \Sigma = E \big \{ \frac{\phi^2 ( \alpha G + \beta T + {X} W \gamma+ X \theta )}{\Phi(\alpha G + \beta T + {X} W \gamma+ X \theta) \Phi(-(\alpha G + \beta T + {X} W \gamma+ X \theta))} ( G , T, {X} W, X)'( G , T, {X} W, X)  \big \} \]
By the delta method we have the following:
\begin{equation}\label{eq7}
    \sqrt{n} (\tau - \Hat{\tau}) \xrightarrow{d} N (0, V)
\end{equation}
with,
\[ V = \Omega(\kappa, X) \Sigma \Omega(\kappa, X)^T, \]
and 
\[ \Omega(\kappa, X) = \frac{\partial \tau (G=1, T=1, X)}{\partial \kappa^T} \]
$\hat{V} = \Omega(\hat{\kappa}, X) \Hat{\Sigma} \Omega(\hat{\kappa}, X)^T$ is a consistent estimator of $V$.

Next, t-statistics and p-values are computed from equation \ref{eq7}. In order to deal with the possibly high false rejection rate due the multiple hypothesis testing, I adjust p-values using Benjamini-Hochberg procedure (BH(q)). The procedure aims to control for the False Discovery Rate.

\section{Estimation Results}
In this section, I present results from the estimation strategies I considered in Section 5. For the nonlinear DiD method I estimate the treatment effect on households with daughters in rural areas by using households with sons in rural areas as the control group.
In subsection 6.1, I present estimates of heterogeneous treatment effect for the cross section study  with household and  age fixed effect. In subsection 6.2, I  present estimates from the nonlinear DiD estimation.

\subsection{Cross-Section Analysis with Household and Age Fixed-Effect}
The estimation coefficients of equations (\ref{eq:2}) are presented in Table \ref{tab2}. I can derive three general conclusions from Table \ref{tab2}. First, as expected a household with educated head of household has in average larger proportion of educated daughters compared to household with non-educated head of household. Second, a household with low wealth index has on average lower proportion of educated daughters. Finally, households with a woman as the head of household have, on average, higher proportion of educated daughters. Concerning the impact of the free elementary education policy on the proportion of educated daughters in a household, I observe a heterogeneous effect as shown by interaction coefficients in table \ref{tab2}. Indeed, households with non-educated head of household and households  with low wealth index respond more to the policy, while households with more children respond less compared to household with fewer children.

\subsection{Nonlinear Difference in Differences Estimation Results}
\begin{figure}[H]
 \centering
 \begin{minipage}[b]{0.45\linewidth}
   \includegraphics [width=8cm]{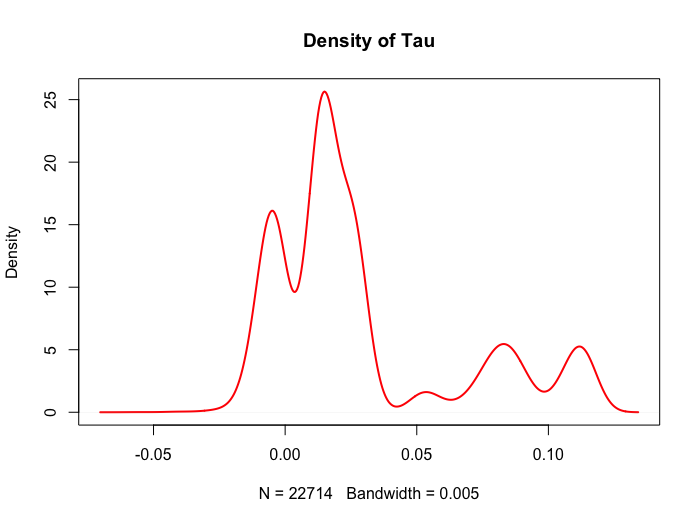}
 \end{minipage}
 \quad
 \begin{minipage}[b]{0.45\linewidth}
    \includegraphics [width=8cm]{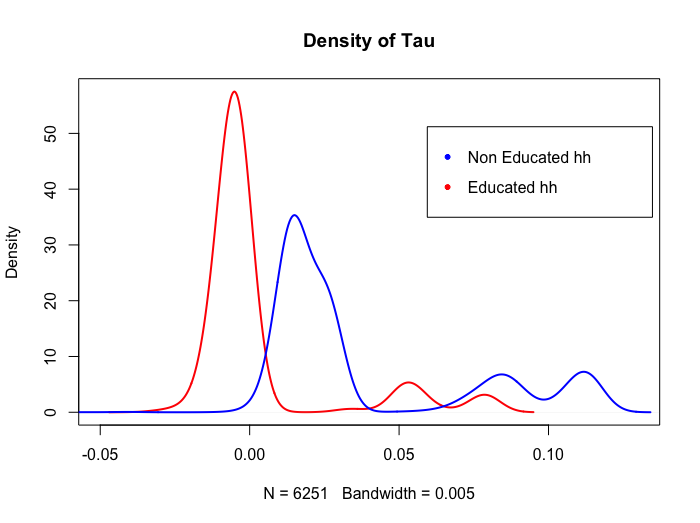}
\end{minipage}

 \quad
 \begin{minipage}[b]{0.45\linewidth}
    \includegraphics [width=8cm]{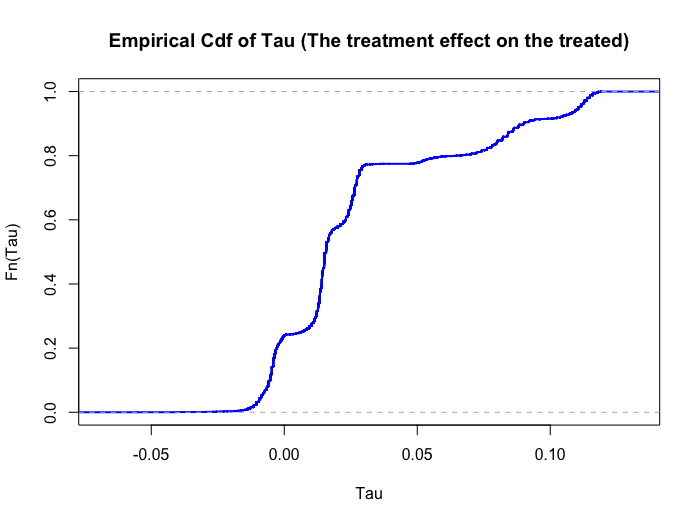}
\end{minipage}
 \quad
 \begin{minipage}[b]{0.45\linewidth}
    \includegraphics [width=8cm]{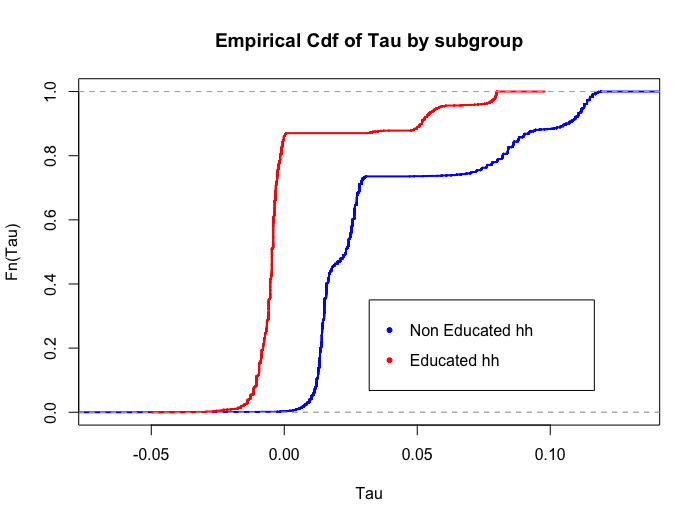}
\end{minipage}
  \caption{Density and CDF plot of the treatment effect on the treated}
 \label{fig8}
\end{figure}

The density and empirical cdf plots of the estimated treatment effect on the treated using nonlinear DiD estimation strategy are presented in Figure \ref{fig8}. With the nonlinear DiD identification strategy, I obtained the treatment effect  for each individual in the treatment group -- household with daughters aged between 13 and 18 years old and located in rural areas. Figure \ref{fig8} indicates that the majority of households responds positively to the free elementary education for girls. However households with non-educated head of household respond more as shown both by the shift to the right of the density of the treatment effect in panel 2 of Figure \ref{fig8} and the first order stochastic dominance observed in the last panel.

\begin{figure}[H]
\centering
 \begin{minipage}[b]{0.45\linewidth}
   \includegraphics [width=9cm]{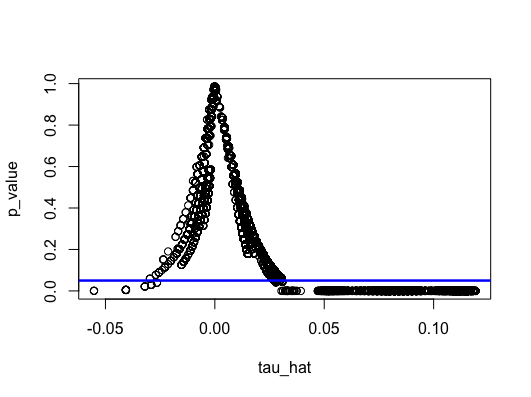}
    (a) p-values
   \label{f1}
 \end{minipage}
 \quad
 \begin{minipage}[b]{0.45\linewidth}
    \includegraphics [width=9cm]{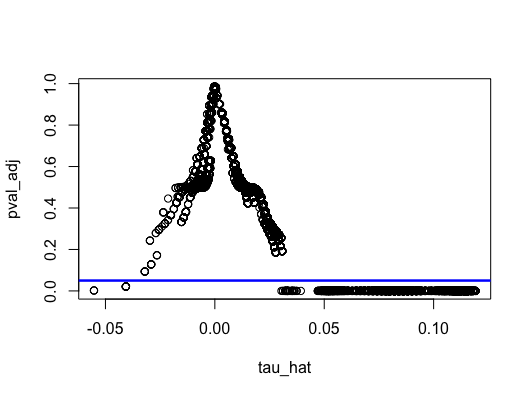}
    (b) adjusted p-values
  \label{f2}
\end{minipage}
    \caption{p-values as function  of the estimated treatment effect on the treated.}
    \label{fig9}
\end{figure}

Next, I investigate the statistical significance of the estimates of the treatment effect on the treated. Figure \ref{fig9} plots the p-values and adjusted p-values as function of the treatment effect. It indicates that treatment effect smaller  than 0.0028 are not statistically significant.  The average treatment effect on the treated--- for individuals with statistically significant treatment effect--- is $8.70\%$. The effect on the treated is $9.21\%$ and $5.89\%$ respectively for non-educated and educated households.

%The conclusion from Figure \ref{fig7} still holds.

\subsection{Constant Time and Group Effect Assumption}
I observe that the waive of tuition fees payment in public primary school in rural Benin has led to an overall increase in the probability that parents enroll their daughter in primary school with non-educated parents responding more than their educated counterparts. The key identification assumption is that the difference in the unobservable variables that govern parents with daughters and parents with sons education decision is constant across time. If this assumption does not hold, then it is possible that over time households with daughters are more aware of the benefits of education and choose to enroll more girls than households with sons. In this case, our estimate of the treatment effect is biased. Failure of the identification assumption also implies the possibility that over time non-educated parents enroll more girls in school than educated parents. In that case, our estimate of heterogeneous treatment effect is biased as well. 

In this section, I consider an approach to check the constant difference between groups \footnote{Treatment-- households with daughters-- and control-- households with sons--.} across time in the unobservables that determine education decision.
For that purpose I examine only observations on children older than 28 years old. And I consider individuals younger than 35 years old as post treatment observations while children older than 34 years old are pre-treatment observations. I then estimate a treatment effect on treated using the same procedure as in section 5.2. Given that children older than 28 years old but younger than 35 years old have not been exposed to any real education policy/program, I should not get any significant estimate of treatment effect if our identification assumption holds. That is exactly what the non significance of the $\gamma$s from equation \ref{eq6} suggests. This result is robust to the threshold for pre- and post-treatment observations (see Appendix Table \ref{table:5}).

\subsection{Conditional Treatment Effect}
\begin{figure}[H]
 \centering
   \includegraphics [width=11cm]{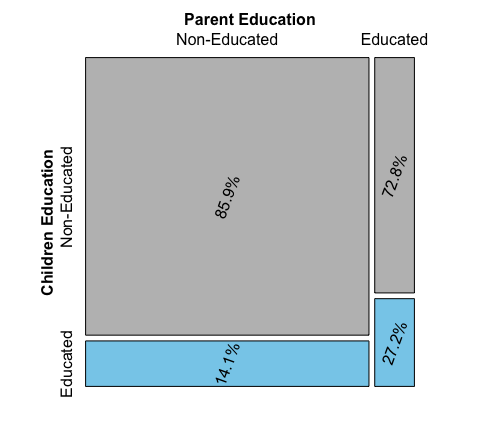}
 \caption{ Education of daughters in rural as function of the education of the head of their households before the policy for sub-sample of households with at least one non-educated child.}
 \label{fig12}\end{figure}

\begin{figure}[H]
 \centering
 \begin{minipage}[b]{0.45\linewidth}
   \includegraphics [width=8.5cm]{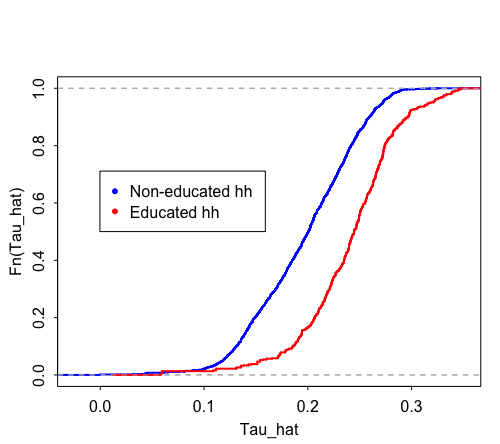}
    %(a) By education of the head of the household
   \label{f13_1}
 \end{minipage}
 \quad
 \begin{minipage}[b]{0.45\linewidth}
    \includegraphics [width=8.5cm]{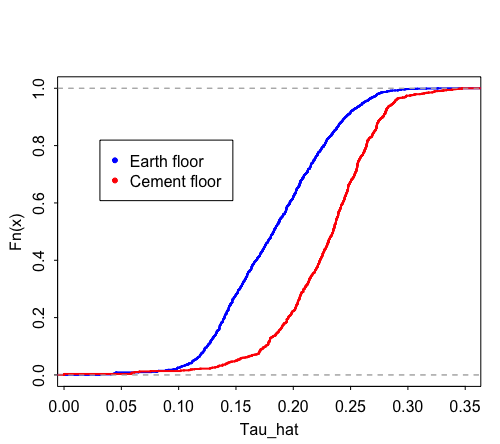}
    %(b) By Household Wealth Index
  \label{f13_2}
\end{minipage}
 \caption{Empirical cdf of the estimated treatment effect for households with at least one $>$ 18 years old non-educated child}
 \label{fig13}
\end{figure}

As argued by the Maximally Maintained Inequality (MMI) hypothesis, my results in section 6 is explained by the fact that the enrollment rate of advantaged children--- children from educated households--- is close to saturation ($\approx 76\%$) as shown in panel (b) of Figure \ref{fig5}. In this section I investigate situation where this enrolment rate is small, by considering the heterogeneous treatment effect conditional on households with at least one non-educated child in pre-treatment periods. This guarantee that educated parents are not educating their child with probability close to 1 in the absence of any education policy. Specifically I estimate the treatment effect for households with at least one non-educated child of 19 years old or older. For this specific sub-sample of the population the enrolment rate of children from educated households ($\approx 27\%$) is quite far from saturation as shown in Figure \ref{fig12}. Consequently, in accordance with the MMI hypothesis, educated households households uniformly benefit more from the education policy as suggested by Figure \ref{fig13}. Similarly,  high HWI households uniformly benefit more from the education policy.

\section{Households' Education decision problem}
Households' preference for education is reflected by the difference in education decision which is function of the differential education level of parents. However, financial resources available also affect education decision. In this section, I consider households' education decision problem and analyze how budget constraint and preference interact when financial cost of education is eliminated. The household decision problem I consider here is based on the model in \cite{ashraf2020bride}. The setup of households' education decision is the following:
\begin{enumerate}
    \item There are multiple households, where $\eta$ is the proportion of households with educated head of household. Let $X_i$ be a vector of household $i$'s specific characteristics with include household wealth index (HWI) and education of the head of household.

    \item Suppose that each household has a value $\beta_i$\footnote{$\beta_i$ can be interpreted as households intrinsic value for education net opportunity cost.} for education. Given that non-educated households tend to have high opportunity cost and low intrinsic value for education I have the following distribution assumptions on $\beta_i$. For educated households I assume that $\beta_i \sim G^1(.)$ and for non-educated households I assume $\beta_i \sim G^0(.)$, where $G^1(.)$ first order stochastically dominates $G^0(.)$. 
    
    \item Next I assume that the wealth of educated households and non-educated households are drawn from distributions $F^1(.)$ and $F^0(.)$ respectively. Where, $F^1$ first order stochastically dominates $F^0$ since educated households tend to have higher wealth compare to their non-educated counterparts. 
    
    \item Suppose that each household has $n_d$ daughters, but the education decisions of daughters $t$ and $t'$ in household $i$ are the same. So, I consider the education decision of one daughter from each household. Let the utility of household $i$ be $U_i = c_i + \beta_i E_i$ and its education decision problem is as follows:
\end{enumerate}

\begin{equation}
    \max_{E_i = \{0, 1\}, c_i}  c_i + \beta_i E_i 
\end{equation}
\[s.t \text{ } c_i + k E_i \leq y_i ,\]

where $y_i$ is the income of household $i$, $E_i$ is the education decision of household $i$, and $k$ is the financial cost of education.\\
If $y_i < k$, then household $i$ can not afford education and choose $E_i = 0$ as result. On the other hand if $\beta_i<0$, education of a child brings negative marginal utility, therefore household $i$ chooses $E_i = 0$.
The utilities of household $i$ when $E_i = 0$  and $E_i = 1$ if $y_i \geq k $ are $y_i$ and $y_i - k + \beta_i$. Therefore household $i$ chooses $E_i = 1$ if and only if $y_i \geq k$ and $ \beta_i \geq k$. This implies that 
\[ Pr[E_i = 1] = Pr( y_i \geq k \text{ and }  \beta_i \geq k) \]

Assuming that $y_i$ and $\beta_i$ are independent\footnote{ In order word I assume that the only channel through which income affect education decision is affordability. This assumption can be justified by the fact that high income but non-educated parents have lower proportion (53\%) of educated daughters compared to low income but educated parents (68\%) for pre-treatment observations. If high income parents had higher $\beta$ compared to low income, not because they also tend to more educated, then non-educated high income household would at least educated daughters as educated low income households. Which is not the case.}, I have the following:

\begin{equation}
    Pr[E_i = 1] = Pr(y_i \geq k ) Pr(\beta_i \geq k)
\end{equation}

When the financial cost of education is eliminated, ie $k = 0$, I have:
\begin{equation}
    Pr[E_i = 1] = Pr(\beta_i \geq 0)
\end{equation}

Considering equations (8) and (9) for educated and non-educated households respectively I have the following expression for the effect of $k= 0$ for each one of them.
\begin{equation}
    \tau^0  = (G^0(k) - G^0(0)) + F^0( k ) (1 -G^0(k)),
\end{equation}
and
\begin{equation}
    \tau^1  = (G^1(k) - G^1(0)) + F^1( k ) (1 -G^1(k)),
\end{equation}
The expressions of $\tau_0$ and $\tau_1$ have two components:
    \begin{enumerate}
        \item The first component $G(k) -G(0)$ is the effect on  probability to choose $E_i = 1$ from free education which allows household with $0 < \beta < k$ to choose $E=1$ from preference perspective.
        \item The second component $F(k)(1-G(k))$ is the combined effect of budget constraint (-) and  preference (+) on the probability of choosing $E_i = 1$ when financial cost of education is not zero.
        \end{enumerate}
Given that $G^1(.)$ first order stochastically dominates $G^0(.)$, 
\[ G^0(k) - G^0(0) > G^1(k) - G^1(0) \]

However, 
\[ F^1( k ) < F^0 ( k) \text{ while } (1 -G^1(k)) > (1 -G^0(k))\]

Therefore, I have the following cases. 
\begin{enumerate}
    \item If disadvantage for non-educated parents from income limitation dominates the advantage for educated parents from preference, ie 
    \[F^0( k ) (1 -G^0(k)) > F^1( k) (1 -G^1(k)),\] then non-educated households benefit more from free education policy.
    \item Otherwise, ie 
    \[F^0( k) (1 -G^0(k)) < F^1( k ) (1 -G^1(k)), \] then non-educated households benefit more from free education policy if G(.), F(.) and $k$ are such that:
    \begin{equation}
        F^1( k ) (1 -G^1(k)) - F^0( k ) (1 -G^0(k)) < (G^0(k) - G^0(0)) - (G^1(k) - G^1(0))
    \end{equation}
     Otherwise educated households benefit more.   
\end{enumerate}

\subsection{Results with Parametric Form for F(.) and G(.)}
I assume the following parametric form for $F(.)$ and $G(.)$.
\begin{itemize}
    \item Assume $y_i^0 = \alpha_i^0 + \varepsilon_i^0$, with 
    $\alpha_i^0 \sim \text{Bernoulli}(p^0)$ and $\varepsilon_i^0 \sim N(0, 1)$. Similarly, $y_i^1 = \alpha_i^1 + \varepsilon_i^1$, with  $\alpha_i^1 \sim \text{Bernoulli}(p^1)$ and $\varepsilon_i^1 \sim N(0.1, 1)$, where $p^1 > p^0$. $\varepsilon_i$s are iid.
    \item Assume $\beta_i^0 \sim N(0, 1)$ and $\beta_i^1 \sim N(\mu,1)$
    \item For simplicity assume that $k = 0.1$ and $p^0 = 0.5$.
\end{itemize}

\begin{figure}[H]
 \centering
 \begin{minipage}[b]{0.45\linewidth}
   \includegraphics [width=7.5cm]{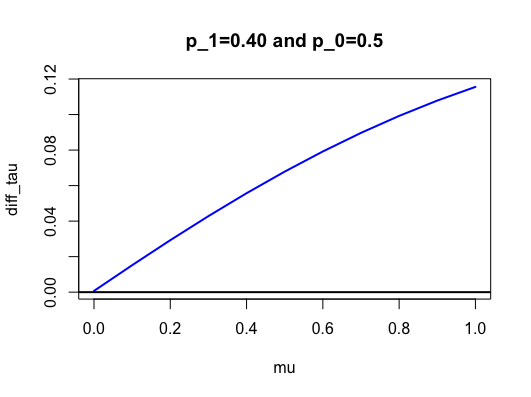}
 \end{minipage}
 \quad
 \begin{minipage}[b]{0.45\linewidth}
    \includegraphics [width=7.5cm]{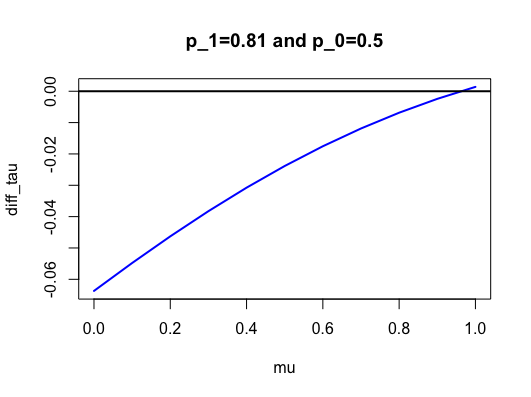}
\end{minipage}

 \quad
 \begin{minipage}[b]{0.45\linewidth}
    \includegraphics [width=7.5cm]{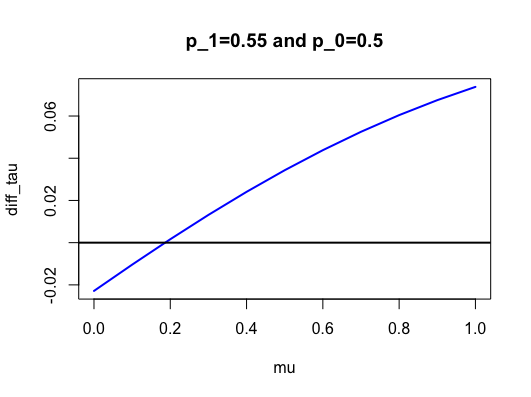}
\end{minipage}
 \quad
 \begin{minipage}[b]{0.45\linewidth}
    \includegraphics [width=7.5cm]{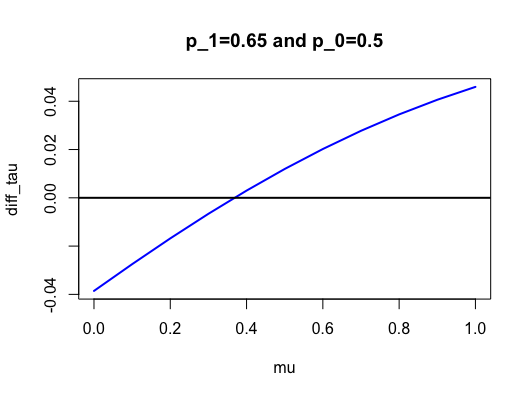}
\end{minipage}
  \caption{Plot of $f(\mu, p^1)$}
 \label{2}
\end{figure}

The function $f(\mu, p^1) = \tau^1 - \tau^0$ plays a crucial role in determining the relative benefits of free elementary education for educated and non-educated households. The direction of $f(\mu, p^1)$ holds the key to this differentiation, where $p^1$ and $\mu$ symbolize parameters representing the income and preference advantages of educated households over their non-educated counterparts. Visualized in Figure \ref{2}, the plots of $g(\mu) = f(\mu,.)$ illuminate this relationship across varying $p^1$ values. Notably, when $p^1$ is sufficiently high (e.g., $p^1 > 0.8$), $f(\mu, p^1)$ takes on a negative stance for all $\mu$ values. Conversely, in scenarios of low $p^1$ (e.g., $p^1 < 0.4$), $f(\mu, p^1)$ has a positive value for all $\mu$. This intriguing pattern suggests that when the income advantage of educated households is significant (resp. minimal) compared to their non-educated counterparts, reductions in educational financial burdens prompt more pronounced shifts in the education decisions of non-educated (resp. educated) households.
However, the analysis takes a nuanced turn for $0.4 \leq p^1 < 0.8$, where $f(\mu, p^1)$ exhibits negativity for lower $\mu$ values. In essence, as the income advantage of educated households expands, a prerequisite for eliciting greater responses to education policy from non-educated households is a proportionally reduced preference advantage held by the educated households. This insight underscores the intricate interplay between income disparities, preference dynamics, and the policy's impact on educational choices.

\section{Conclusion}

In this paper, I delve into the assessment of the impact of a reform that introduces free elementary education for girls in the rural regions of Benin. To achieve this, I establish a pseudo panel data framework wherein I consider each household as an observational unit. This framework allows me to track the educational choices made for every child within a household. Using households with male children in rural areas as control group, I employ a nonlinear difference in differences approach to estimate the effect of the policy.

The outcomes of the analysis reveal a noteworthy trend: households headed by individuals with limited or no formal education gain more pronounced benefit from the reform compared to households with educated heads. Furthermore, upon closer examination of households that had at least one uneducated daughter before the policy implementation, it becomes evident that educated households exhibit a stronger response compared to their non-educated counterparts. This empirical confirmation aligns with the Maximally Maintained Inequality hypothesis.

The paper also delves into a conceptual framework centered around household decision-making concerning education. Through this lens, I explore how the interplay between a household's preference towards education and its budgetary limitations contributes to the observed empirical findings. The conclusions drawn from this analysis underscore a significant insight: as the income disparity between educated and non-educated households widens, the effectiveness of education policies in prompting a response from non-educated households is contingent upon the relative decrease in preference advantage held by educated households.

\newpage
\section*{Figures and Tables}
% Figure of HWI

 % Histogram of elementary education

 % Structure of the population

% No spillover effect
 
% education across gender and cohort

% structural break

% Nonlinear DiD results

% p-values

%Conditional ATT
 
% Structural model

% OLS results

\begin{table}[H]
    \caption{Estimation coefficients of equations (2) \tablefootnote{standard errors are reported in parenthesis. Significance. codes:  0 ‘***’ 0.001 ‘**’ 0.01 ‘*’ 0.05 ‘.’ 0.1 ‘ ’ 1.}}
    \begin{minipage}{16cm}
     
 \centering\begin{tabular}{p{7cm}  p{3cm} p{3cm} p{3cm}}
& & &\\
\hline
\hline
&\multicolumn{3}{c}{Coefficient}\\
& (1) &(2) & (3) \\
\hline
Educated hh & $ 0.2784^{***}$ & $0.2896^{***}$&  $0.2771^{***}$\\
 & (0.0130) &(0.0233)& (0.0236)\\
Low HWI & $-0.1684^{***}$ &$-0.1778^{***}$& $-0.1743^{***}$\\
 & (0.0123)& (0.0219)& (0.0223) \\
N & $-0.0072^{***}$  & -0.0032& $-0.0069^{*}$\\
& (0.0009) & (0.0025)& (0.0027)\\
Muslim & $-0.0842^{***}$  & $-0.0401^.$ & -0.0365 \\
& (0.0136) & (0.0225)& (0.0227)\\  
Christian & $0.0983^{***}$  & $0.1167^{***}$& $0.1135^{***}$\\
 & (0.0116)& (0.0175)& (0.0176)\\  
Female hh & $0.0715^{***}$ & $0.0649^{**}$& $0.0624^{**}$\\
 & (0.0111) &(0.0205)& (0.0207)\\  
  $R$ & -0.01623 &  0.0298& -0.0221 \\
 & (0.0300) & (0.0234)&  (0.0301)\\  
$R \times \text{Low HWI}$ & $0.0393^{**}$ & $0.0400^{*}$& $0.0389^{*}$ \\
&  (0.0158) & (0.0160)& (0.0162) \\  
$R \times N$ & $-0.0063^{*}$ & $-0.0072^{***}$ & $-0.0062^{**}$  \\
& (0.0023) &(0.0022)& (0.0023)\\
$R \times \text{Non-Educated hh}$ & $0.0279^{.}$ & $0.0448^{**}$& \textbf{$0.0346^{*}$}\\
& (0.0161) & (0.0161)& (0.0163)\\  
Average age fixed effect  & Yes& No \footnote{Control only for within household average age of girls.}& Yes\\
Interaction between covariates\footnote{Interaction between head of household's education, HWI, number of children, gender of hh and religion.} &No& Yes& Yes\\
Number of observations  & 23472 &23472&23472 \\
$R^2$ & 0.697  &0.2253& 0.6979\\  
\hline
\label{tab2}
\end{tabular}
    \end{minipage}
\end{table}

\begin{table}[!th]
\caption{Two way table of education of daughters and head of households}

\centering
\begin{tabular}{ p{5cm } p{3cm} p{3cm} p{2cm}}
& &&\\
\hline
\hline
& Educated hh &Non-educated hh &Total \\
\hline

 Educated daughter &17.80& 28.71& 46.51 \\

 Non-educated daughter & 5.67 & 47.81 & 53.49\\

 Total &23.47& 76.52 & 100 \\
\hline
\hline
\label{tab3}
\end{tabular}
\end{table}

\newpage

\bibliographystyle{plainnat}
\bibliography{References}

\newpage
\section{Appendix}
\subsection{Intergenerational Educational Persistence}

\begin{table}[!th]
\caption{Variance-Coviance Matrix of Wealth and Education Variables }

\centering
\begin{tabular}{ p{6cm } p{4cm} p{4cm} p{3cm}}
& &&\\
\hline
\hline
& Education of hh &Education of children &HWI \\
\hline

 Education of hh &1& 0.47 &0.57\\

 Education of children & 0.47 & 1 & 0.48\\

 HWI &0.57& 0.48& 1 \\
\hline
\hline
\label{table:3}
\end{tabular}
\end{table}

\begin{table}[!th]
\caption{Intergenerational mobility indicators in rural areas}
\centering
\begin{tabular}{ p{6cm } p{3cm} p{3cm}}
& &\\
\hline
\hline
& Sons & Daughters \\
Mobility indicator & 62.03\% & 59.82\%\\

Ascending mobility & 53.37\% &47.38\% \\

Descending mobility & 8.67\% &12.44\% \\
\hline
\hline
\label{table:4}
\end{tabular}
\end{table}

\begin{table}[!th]
\caption{Placebo Analysis}
\centering
\begin{tabular}{ p{5cm } p{4cm} p{4cm} p{4cm}}
& & &\\
\hline
\hline
 &$T = 1\{ Age < 35\}$ & $T = 1\{ Age < 45\}$ & $T = 1\{ Age < 55\}$ \\
 & & &\\
$1\{\text{Female } \& \text{ Rural} \} \times T$ & -0.11 (0.14)  &  0.12  (0.16)  & 0.18   (0.27)\\
& & &\\
$1\{ \text{Female } \& \text{ Rural} \} \times T \times 1\{\text{Educ hh = 0}\}$& 0.14 (0.10) &0.08   (0.09) & 0.05   (0.09)\\

$1\{ \text{Female } \& \text{ Rural} \} \times T \times 1\{\text{Cement floor = 0}\}$&-0.07 (0.08) & -0.06   (0.07)& -0.05   (0.07)\\

Other Covariates & Yes&Yes &Yes\\
\hline
\hline
\label{table:5}
\end{tabular}
\end{table}

\begin{table}[H]
\caption{Migration status (previous residence) of the sample of children by gender in rural areas}
\centering
\begin{tabular}{ p{5cm } p{3cm} p{3cm} p{3cm}}
& & & \\
\hline
\hline
& Sons & Daughters & Overall \\
& & & \\
Same major, same minor administrative unit &  90.49   &  88.97 &  89.87 \\
\hline
\hline
\label{table:6}
\end{tabular}
\end{table}
Major administrative units are departments and minor administrative units are communes. Therefore within commune migration is not accounted for here.

\subsection{Heterogeneous treatment effect across HWI and number of children}

  \begin{figure}[H]
 \centering
 \begin{minipage}[b]{0.45\linewidth}
   \includegraphics [width=9cm]{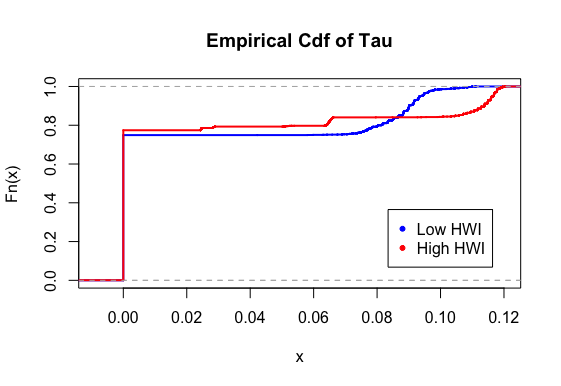}
   (a) Household wealth index
 \end{minipage}
 \quad
 \begin{minipage}[b]{0.45\linewidth}
    \includegraphics [width=9cm]{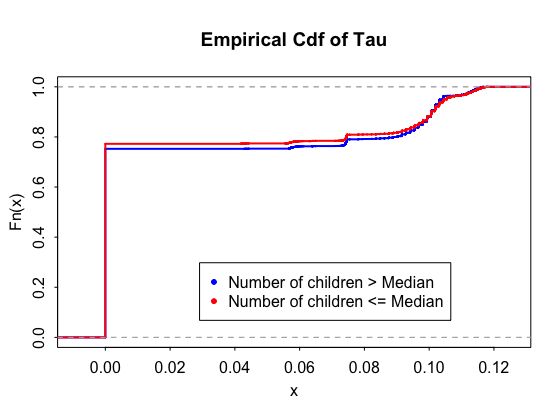}
    (b) Number of children
\end{minipage}

  \caption{Empirical cdf of tau by sub-group}
 \label{fig14}
\end{figure}

\subsection*{Proof of equation (4)}

Define latent potential outcomes $Y^{0 \star}$ and $Y^{1 \star}$  respectively as:
\[Y^{0\star}_{it} = \alpha G_{it} + \beta T_{it}   + \theta X_i +  \varepsilon_{it}, \]
and 
\[Y^{1\star}_{it} = \alpha G_{it} + \beta T_{it} + \gamma ({X}_i W_{it})   + \theta X_i +  \varepsilon_{it}, \]
such that:
\[E[Y^0|T,G,X] = \Phi(\alpha G + \beta T + X \theta),\]
and 
\[E[Y^1|T,G,X] = \Phi(\alpha G + \beta T +  {X} W \gamma+ X \theta)\]

Therefore 
\begin{equation*}\label{eq:6}
    \tau(G=1, T=1, X) = \Phi(\alpha + \beta +  {X} W \gamma+ X \theta) - \Phi(\alpha + \beta + X \theta)
\end{equation*}

\end{document}